\documentclass[aps,prl,twocolumn,groupedaddress,showpacs]{revtex4-1}
\usepackage{graphicx}
\usepackage{psfrag}
\usepackage{amsmath}
\def\be{\begin{equation}}
\def\ee{\end{equation}}

\begin{document}
\date{\today}

\title{Topology of Collisionless Relaxation }
\author{Renato Pakter}
\author{Yan Levin}
\affiliation{Instituto de F\'{\i}sica, Universidade Federal do Rio Grande do Sul, Caixa Postal 15051, CEP 91501-970,
Porto Alegre, RS, Brazil}

\begin{abstract}

Using extensive molecular dynamics simulations we explore the fine-grained phase space
structure of systems with long-range interactions. We find that if the initial phase space 
particle distribution
has no holes, the final stationary distribution will also contain 
a compact simply connected region. The microscopic holes created by the filamentation of the initial distribution function
are always restricted to the outer regions of the phase space.
In general, for complex multi-level distributions it is very difficult 
to {\it a priori} predict the
final stationary state without solving the full dynamical evolution.  However, we show that for multi-level
initial distributions satisfying a generalized virial condition, it is possible
to predict the particle distribution in the final stationary state using Casimir invariants of the Vlasov dynamics.

\end{abstract}


\maketitle

Statistical mechanics of systems in which particles interact through long-range (LR) forces remains an outstanding challenge\cite{Campa09}.  Unlike short-range systems,
which relax to thermodynamic equilibrium through  binary collisions, systems with LR interactions become trapped
in the quasi-stationary states (qSS) the life time of which diverges with the number of particles~\cite{LaRa99,YaBa04,AnFa07,JaBo07}.  The fundamental difficulty in studying the qSS 
is that these states explicitly depend on the initial particle distribution. 
Furthermore, the
lack of ergodicity intrinsic to LR systems prevents application of the standard
equilibrium statistical mechanics which has proven to be so powerful for
describing the many-body systems with finite-range forces~\cite{BeTe12}.  
In spite of these
difficulties, it has been recently observed that there is a significant degree of universality in the process of collisionless relaxation. Many different systems ranging 
from plasmas~\cite{LePa08} to gravitational clusters~\cite{TeLe10,JoWo11} 
have been found to relax to qSS with a characteristic core-halo structure. 

In  thermodynamic limit, LR systems are
collisionless  -- particles move under the action of the mean-field potential produced by all the other particles. In general, the mean-field potential has a complex
dynamics, characterized by quasi-periodic oscillations. It is possible, therefore, for some particle to enter in resonance with the oscillations and to gain large amounts of energy at the expense of the collective motion. This process, known in plasma physics as the Landau damping~\cite{La46}, diminishes the amplitude of oscillations
and leads to the formation of a tenuous halo of highly energetic particles which  surrounds the high density core.  When all the oscillations die out a qSS will be born.  The
qSS is characterized by a broken ergodicity  --- since the only mechanism through which particles can gain or loose energy
is the Landau damping, once the mean-field potential becomes stationary, the dynamics of each particle  becomes integrable (for spherically symmetric systems)~\cite{BeTe12}.  When this happens there is no longer a mechanism through which highly energetic particles of the halo can  equilibrate with the particles of the core. 

The relaxation to qSS is very similar to the process of evaporative cooling.  As some particles enter in resonance with the collective oscillations they gain energy, while cooling down the core region. The Hamiltonian dynamics of LR systems is governed by the collisionless Boltzmann (Vlasov) equation~\cite{Br77}.  This equation requires that
the one-particle distribution function evolves in the phase space as the density of an incompressible fluid.
This means that the core region can not collapse to the minimum of the potential energy-- since this would violate the incompressibility requirement imposed by the Vlasov flow -- instead the maximum phase space density can not exceed that of the initial  distribution.  For one-level initial distributions this observation allows us to predict the qSS without having to explicitly solve the Vlasov equation or perform the MD simulations. When all the oscillations have died out, the particles in the core should occupy all the low-energy states up to the maximum phase space density permitted by the original one-level waterbag distribution.  The particle distribution inside
the core will then be the same as that of a fully degenerate Fermi 
gas~\cite{LePa08,TeLe10,PaLe11}.  
On the other hand the particles
in the halo will be approximately uniformly distributed up to the maximum energy corresponding to the location of the parametric resonance.  The  Fermi energy and the phase space density of the halo particles can then be obtained from the requirement of the conservation of the norm and of the total energy of the system.  The theory has been found to be extremely successful, allowing us to predict the distribution functions of confined plasmas~\cite{LePa08}, 1d and 2d gravitational systems~\cite{TeLe10}, the HMF model~\cite{PaLe11}, etc. The question that we would like to address in this Letter  is how to extend the theory described above to more complex initial particle distributions~\cite{TePa09,AsFa12,CaCh12}.

To be specific we will study a class of distributions 
which are compact with a simply connected support (no-holes).
The distributions have $L$ different phase space density levels, see Fig.1a.  
To demonstrate the theory we will 
use a paradigmatic model of a system with LR forces known as the 
Hamiltonian Mean-Field (HMF )model~\cite{Campa09}.
The HMF model describes $N$ particles that are constrained to move on a 
circle of radius one.  The dynamics 
is governed by the Hamiltonian
\be
H=\sum_{i=1}^N {p_i^2\over 2}+{1\over 2 N}\sum _{i,j=1}^N [1-\cos(\theta_i-\theta_j)],
\ee
where the angle $\theta_i$ is the position of  $i$'th particle
and $p_i$ is its conjugate momentum~\cite{AnCa07,AnFa07}.
The {\it macroscopic} behavior of the system is characterized by the magnetization vector
${\bf M}=(M_x,M_y)$, where $M_x\equiv \langle \cos\theta \rangle$, 
$M_y\equiv \langle \sin\theta \rangle$, and $\langle\cdots\rangle$ stands for the average
over all  the particles.  The Hamilton's equations of motion for each 
particle reduce to  
\be
\ddot\theta_i=-M_x(t)\sin\theta_i(t)+M_y(t)\cos \theta_i(t).
\label{evol}
\ee
Since the Hamiltonian does not have an explicit time dependence, the average energy per particle,
\be
u={H\over N}={\langle p^2\rangle\over 2}+{1-M(t)^2\over 2}\,,
\label{energyu}
\ee 
is conserved. For symmetric distributions $(\theta \rightarrow -\theta)$
$M_y=0$ throughout the evolution, so that the macroscopic dynamics is
completely determined by $M_x(t)$, which for simplicity we will write as $M(t)$~\cite{PaLe11,RoAm12}.

Using the MD simulations, we first calculate the one-particle distribution 
function in the qSS, $f(\theta, p)$.  From Jean's theorem, in the
stationary state
$f(\theta, p)=f(\varepsilon)$, 
where $\varepsilon(\theta,p)=p^2/2+1-M\cos\theta$ is the one-particle energy and $M=\langle \cos\theta \rangle$ is the magnetization. 
The initial phase space 
particle distribution is shown in Fig 1a.  It consists of three different
levels with the phase space densities -- from inside to outside --  $\eta_1$,
$\eta_2$, and $\eta_3$.  To obtain the $f(\varepsilon)$ we run the simulation
until the systems has reached a qSS.  We then separate all the particles 
into bins of energy width $ d\varepsilon$.  To calculate the distribution function  
$f(\varepsilon)$
the fraction of the particles in each bin is divided by the density of states 
\be
g(\varepsilon)=\int \delta\left (\varepsilon-p^2/2-1+M\cos\theta \right )dp\,d\theta,
\ee
where $\delta(x)$ is the Dirac delta function and the integral is performed over all 
the phase-space, $-\infty <p<+\infty$,
$-\pi<\theta<\pi$. Using  $\delta[f(x)]=\sum_i \delta(x-x_i)/|f'(x_i)|$,
where $x_i$ are the roots of $f(x)$, the integration in momentum leads to
\be
g(\varepsilon)=2\int_{-\theta_{max}}^{\theta_{max}}{d\theta\over\sqrt{2\left(\varepsilon-1+M\cos\theta\right )}},
\label{ge1}
\ee
where $\theta_{max}=\pi$ if $\varepsilon>1+M$, and $\theta_{max}=\cos ^{-1}[(1-\varepsilon)/M]$ if $\varepsilon<1+M$. 
Without loss of generality we will take $M>0$.
Performing the integral in Eq.~(\ref{ge1}) yields
\be
g(\varepsilon)=\begin{cases} 4M^{-{1\over 2}} \, K(\kappa), & \mbox{if } \kappa\leq 1, \\ 
4(M\kappa)^{-{1\over 2}} \, K(\kappa^{-1}), & \mbox{if } \kappa>1, \end{cases}
\ee
where $\kappa\equiv (\varepsilon-1+M)/2M$, and $K(\kappa)$ is the complete elliptic integral of the first kind.
Note that $\kappa<1$ ($\kappa >1$) corresponds to librating (rotating) orbits.  
The results of these calculations are shown in Fig. 1.
\begin{figure} 
\includegraphics[scale=0.8,width=9cm]{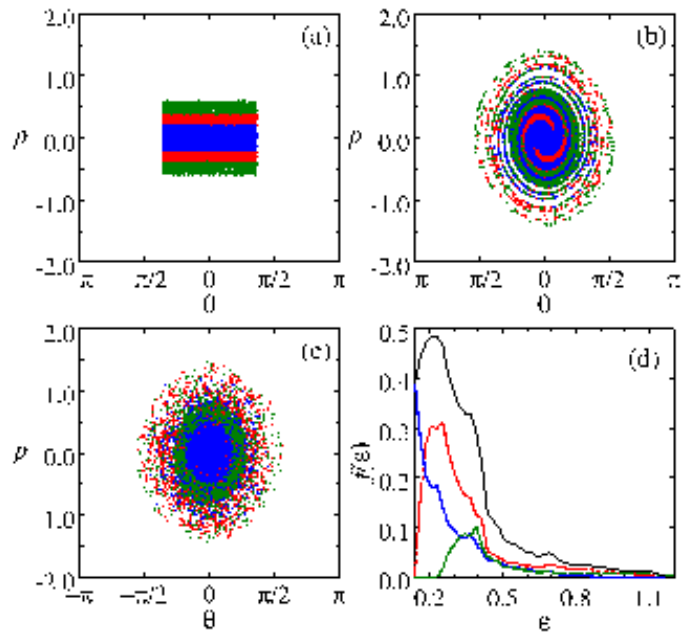}
\caption{(color online). (a) The initial 3 level phase space distribution.  Colors are used to denote
the difference phase space densities. There are a total of 
$N=10^6$ particles distributed over three three level with densities:  
$\eta_1=0.39$ ($0<|p|<p_{max}/3$) in blue, $\eta_2=0.56$ ($p_{max}/3<|p|<2p_{max}/3$) in red, and 
$\eta_3=0.17$ ($2p_{max}/3<|p|<p_{max}$)
in green. The initial magnetization is $M_0=0.8$ $(\theta_{max}=1.131)$ and $p_{max}=0.59$.  Panel (b) shows
a snapshot of phase space at $t=200$, demonstrating the mechanism of mixing through  filamentation.  Panel (c) shows a phase space snapshot at $t=20000$, after the qSS has been established.  Notice the characteristic core-halo structure of the particle distribution.  In panel (d) we plot the total distribution function $f(\epsilon)$ in the qSS (solid black curve) and the partial distribution functions $f_n(\epsilon)$ for each phase space level (dashed curves).}
\label{fig1}
\end{figure}
The panel (a) of Fig. 1 shows the initial particle distribution. 
The stretching and folding, characteristic of collisionless relaxation, appears
as the filamentation of the original phase space, panel (b). 
Fig. 1c provides
a snapshot of the phase space particle distribution in the qSS.  
Notice the appearance of
the characteristic core-halo structure.  In panel (d)
we plot the total particle distribution function $f(\varepsilon)$ 
in the qSS, and the partial distribution functions $f_n(\varepsilon)$ 
for particles which at $t=0$ 
were inside the levels of different phase space densities $\eta_n$.  
The distribution functions have a complicated
structure which, unfortunately, shed very little light on the properties of the qSS. 

The Vlasov dynamics requires that the phase 
space area occupied by each density level
of the initial distribution function must be preserved throughout 
the dynamical evolution~\cite{Ly67}.  To explore this feature we next study the
fraction of the phase space volume at energy $\varepsilon$
occupied by the level $\eta_n$,
$\phi_n(\varepsilon)=f_n(\varepsilon)/\eta_n$.  In Fig. 2 we plot
the $\phi_n(\varepsilon)$ in the qSS for a 3-level 
distribution of Fig. 1.  Once again we see a very complex mixing of the density levels
over the phase space.  Amazingly, however, when all the volume fractions are summed together, a very simple core-halo structure emerges, the solid curve of Fig. 2.  
\begin{figure} 
\includegraphics[scale=0.8,width=9cm]{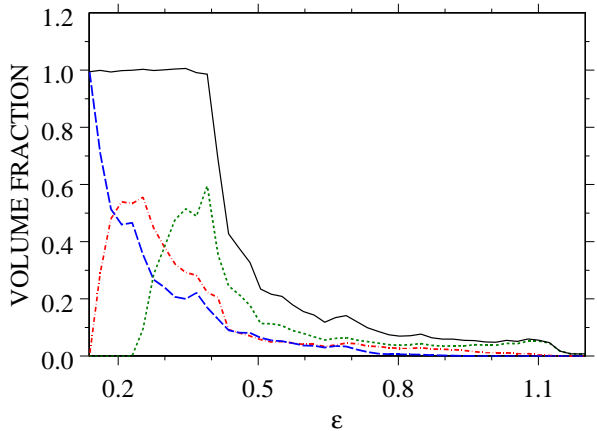}
\caption{(color online). The volume fractions $\phi_n(\varepsilon)$ of the phase space occupied by the 3 density levels
in the final qSS: $\eta_1$ (blue long dashed curve), $\eta_2$ (red dot-dashed curve), 
$\eta_3$ (green short dashed curve).  The black solid curve shows the volume fraction occupied by sum of all three density levels.  Note that in the core region, which extends up to $\varepsilon \approx 0.4$, all the phase-space is fully occupied.  
On the other hand the halo is $90\%$  empty phase space.}
\label{fig2}
\end{figure}
For energies up to $\varepsilon \approx 0.4$, the phase space is completely 
occupied by the density levels of the original distribution function.  There 
are no holes in the central "core" region. All the vacancies are confined
to the outer "halo" region. Indeed, almost $90\%$ of the halo is an empty phase space. 

The simplicity of the core-halo structure suggests that it might be possible to
predict the final qSS {\it a priori}, without having to solve explicitly the
full many-body dynamics.  This, indeed, was the case for the one-level waterbag distributions~\cite{PaLe11},
for which the core was perfectly described by the fully degenerate Fermi-Dirac (FD) distribution, while the halo energy
could be calculated using the theory of parametric resonances.  Unfortunately the situation is much more
complex for many-level systems. 
To get a better feel for the dynamics leading to formation of qSS, in Fig. 3 
we show a schematic evolution of a two-level system from the 
initial to final state.  
The phase space is divided into macrocells of area $dpd\theta$. Each macrocell is then
subdivided into microcells.  The incompressibility of Vlasov dynamics requires that
each microcell is occupied by at most one density level.  In the qSS the central core region does not contain any microscopic holes (white squares) which are all 
confined to the outside halo region.  
Although the evolution leads to a completely degenerate core (with no holes) it can not, in general, be described by the FD statistics. The problem is that the degenerate limit ($T=0$) of the FD statistics requires that the low energy states  must be occupied by the levels with largest value of $\eta_n$. To be in the ground state 
a FD system must be stratified so that for $\varepsilon<\varepsilon_1$ only levels $\eta_a$ should be present, for  $\varepsilon_1< \varepsilon<\varepsilon_2$ there should be only levels $\eta_b$, for  $\varepsilon_2< \varepsilon<\varepsilon_3$ there should be only levels $\eta_c$ etc.,  where $\eta_a>\eta_b>\eta_c>...$ and $\varepsilon_1,\varepsilon_2,\varepsilon_3...$ are the Fermi energies.  This, however, is not the
case, as can be clearly seen from Figs. 2 and 3.  There is a mixture
of different levels inside the core region. 
Thus, the statistics of density levels can not, in general, be obtained {\it a priori} without explicitly solving the full many-body dynamics.  
\begin{figure} 
\includegraphics[scale=0.8,width=9cm]{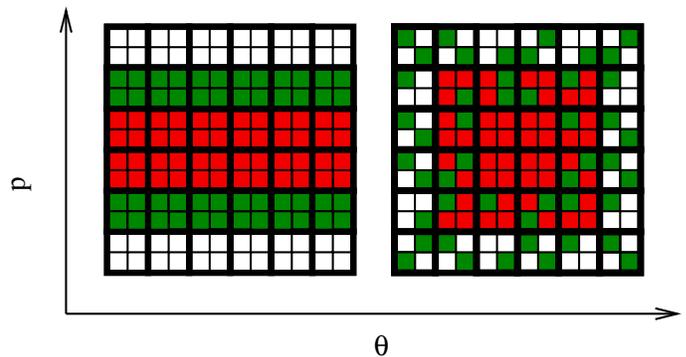}
\caption{(color online). Schematic of phase space evolution for a virialized two-level distribution: panel (a) is the initial condition; panel (b) the fine-grained distribution function in 
the final qSS.  Note that the central core region has no holes (white squares). For a virialized initial condition, the oscillations are suppressed and the resonances are not excited. For such initial distributions the halo is populated only by the "outer" (green) level}
\label{fig3}
\end{figure}  

The complex mixture of different density levels is a consequence of the parametric resonances  produced by the
particle-wave interactions.  If the resonances can be suppressed, the structure of the qSS should be much simpler~\cite{BeTe12}.  The resonances will not be excited if the initial 
particle distribution 
satisfies a generalized virial condition (GVC).  To derive the GVC we require that 
the mean square angle $\langle\theta^2\rangle$ does not vary significantly with time, i.e. that there are no strong envelope oscillations. This will be the case if the two temporal derivatives of $\langle\theta(t)^2\rangle$ vanish, i.e. if
$\langle\theta p\rangle =0$ and 
$\langle p^2\rangle -\langle \cos\theta \rangle \langle\theta \sin \theta \rangle=0$. 
The GVC can be satisfied by adjusting 
the maximum momentum of the particles of each density level in the initial distribution function. 

In the absence of resonances there is no mechanism for the individual particles to gain energy.  
Therefore, the maximum one-particle
energy in the qSS will be the same as the maximum one-particle energy in the initial distribution. For a $L$-level distribution function satisfying the GVC, the mixing will be restricted to the consecutive energy levels $[\varepsilon_i,\varepsilon_{i+1}]$, allowing us to write a simple ansatz solution for the distribution function:
\begin{eqnarray}
&&f_s(\theta,p)=\eta_1\Theta(\varepsilon_1-\varepsilon)+\nonumber\\ && \sum_{i=1}^{L} \left[(1-\chi_i)\eta_{i+1}+\chi_i\eta_{i}\right]
\Theta(\varepsilon_{i+1}-\varepsilon)\Theta(\varepsilon-\varepsilon_{i}),
\label{fs}
\end{eqnarray}
where $\{\varepsilon _i\}$ are the $L+1$ threshold energies that separate
regions of different phase-space density. 
$\chi_i$ is the the fraction of the phase space volume occupied by the level $\eta_i$ in the the phase-space region with energy $[\varepsilon_i,\varepsilon_{i+1}]$. We define $\eta_{L+1}\equiv 0$ and 
$\varepsilon_{L+1}=\varepsilon_{max}=p_{max}^2/2+1-M_0\cos\theta_{max}$, 
which is the
energy of the most energetic particle from the initial condition.
Lack of resonances permits the density level transfers only between the consecutive
energy levels $[\varepsilon_i,\varepsilon_{i+1}]$ and $[\varepsilon_{i+1},\varepsilon_{i+2}]$.
The conservation of the phase space volume occupied by each density level provides us with  
$L$ coupled  equations
\be
V_1+\chi_1 V_2=\nu_1/\eta_1,
\ee
\be
(1-\chi_{i-1})V_i+\chi_i V_{i+1}=\nu_i/\eta_i,\ \ i=2, L,
\ee
where $\nu_i$ is the fraction of particles in the density level $\eta_i$, and
$V_i=\int _{\varepsilon _{i-1}}^{\varepsilon _{i}}g(\varepsilon)\, d\varepsilon$ is the total phase-space volume between 
the energies
$\varepsilon _{i-1}$ and $\varepsilon _{i}$.  The minimum energy is $\varepsilon _{0}= 1-M_s$,  where  $M_s$ is the magnetization of the qSS.      

The phase space volumes transfered from one energy level to the subsequent energy level must be conserved.  This means that all 
the $\chi_i$'s are related
by $\chi_{i-1} V_i=\chi_{i}V_{i+1}$. 
Finally, the conservation of the total 
energy per particle
and the self-consistency requirement for magnetization,
\be
\int \left({p^2\over 2}+{1-M_s\cos\theta\over 2}\right) f_s(\theta,p)\,d\theta dp=u,
\ee
\be
\int \cos\theta \, f_s(\theta,p)\,d\theta dp=M_s,
\ee
give us a total of $2 L+1$ equations to determine $M_s$, and the $L$ threshold energies $\{\varepsilon_i\}$ and 
$\{\chi_i\}$ values. All these equations have to be solved
self-consistently to obtain the distribution function in the qSS, Eq. (\ref{fs}).   
In Fig. 4 we present the solution of these equations for various two-level
distribution functions satisfying the GVC. To compare with MD simulations, we plot both the  position and the velocity distribution functions.
As can be seen the agreement between the theory and the simulations is excellent, without any adjustable parameters.

\begin{figure} 
\includegraphics[scale=0.8,width=9cm]{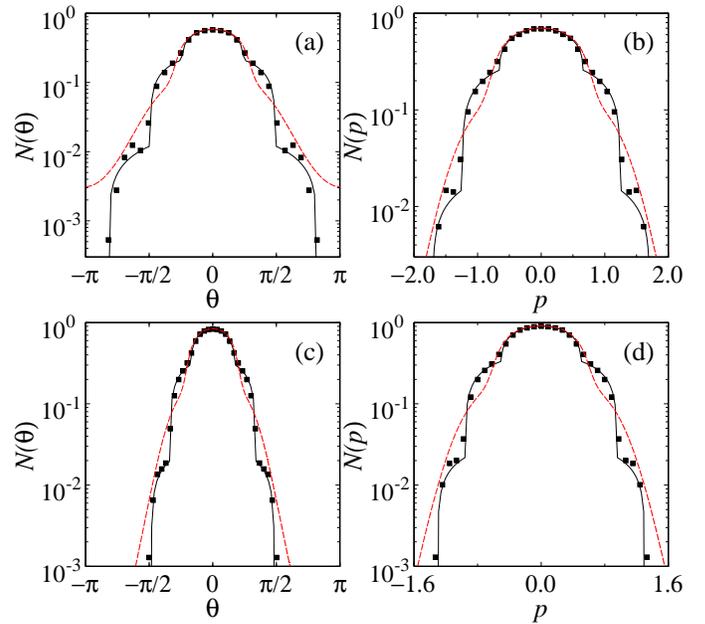}
\caption{The position and the velocity distributions in the qSS state.  The symbols are
the results of MD simulations. The dashed red curves are the predictions of the LB theory, 
while the solid black curves are the prediction of the present theory.  There are no adjustable parameters.  The initial two-level distribution function satisfies the GVC.  We consider two examples:
panels (a) and (b) are the results for the initial distribution
with $\eta_1=0.35$ ($0<|p|<p_{max}/3$), 
$\eta_2=0.058$ ($p_{max}/3<|p|<p_{max}$), $M_0=0.8$, $u=0.33$ ($p_{max}=1.4$, $\theta_{max}=1.131$).
Panels (c) and (d) are the results for the initial distribution
with $\eta_1=0.66$ ($0<|p|<p_{max}/3$), 
$\eta_2=0.11$ ($p_{max}/3<|p|<p_{max}$), $M_0=0.9$, $u=0.18$ ($p_{max}=1.1$,
$\theta_{max}=0.787$).}
\label{fig4}
\end{figure}

It is interesting to compare the predictions of the present theory with the approach based
on the maximization of the coarse-grained Lynden-Bell (LB) entropy~\cite{Ly67}.  Within the LB statistics the distribution function is predicted to be 
\begin{equation}
f(\theta,p)=\sum_{j=n}^L \eta_n \phi_{n}(\theta,p) \;,
\label{fc0}
\end{equation}
where
\begin{equation}
\phi_n(\theta,p)=
\frac{e^{-\beta \eta_n\epsilon(\theta,p)+\alpha_n}}{1+\sum_{i=1}^L e^{-\beta \eta_i \epsilon(\theta,p) +
\alpha_i}}\;,
\label{fc}
\end{equation}
and $\beta$ and $\{\alpha_j\}$ are the Lagrange multipliers used to conserve the total
energy and the phase space volume of each density level.  For one-level waterbag
distributions the LB theory was found to work very well when the initial distribution function
satisfied the GVC.  This, however, is no longer the case for the multilevel distributions.  
Fig. 4, shows that for such distributions the predictions of the LB theory deviate significantly from the 
results of MD simulations. In particular, the LB theory violates the topological 
constraint that the stationary distribution function must have no microscopic holes in the central core region.  
  
We have studied the dynamics of collisionless relaxation of systems with LR forces.
It was found that if the initial phase-space 
particle distribution
has compact simply connected support -- has no holes -- the final stationary distribution will also contain 
a compact region. We find that the microscopic holes created by the filamentation of the initial distribution function
are always restricted to the "outer" regions of the phase space resulting in a
characteristic core-halo structure.

For an arbitrary  initial multi-level distributions it is very difficult
to {\it a priori} predict the
final  qSS without solving explicitly the full many-body dynamics.  An incomplete relaxation and a non-ergodic mixing of the different density levels prevents the
use of standard methods of statistical mechanics.  However, we find that for multi-level
initial distributions satisfying the  GVC it is possible to
{\it a priori} predict the particle distribution in the  qSS without any adjustable
parameters.  The challenge now is to extend the theory to the initial distributions which are not  virialized.    
   
This work was partially supported by the CNPq, FAPERGS, INCT-FCx, and by the US-AFOSR under the grant FA9550-12-1-0438.

\end{document}